# Supercritical Water Gasification: Practical Design Strategies and Operational Challenges for Lab-Scale, Continuous Flow Reactors

*Brian R. Pinkard, David J. Gorman, Kartik Tiwari, Elizabeth G. Rasmussen, John C. Kramlich, Per G. Reinhall, Igor V. Novosselov[1]*

*Mechanical Engineering Department, University of Washington, Seattle, WA 98195*

## Abstract

Optimizing an industrial-scale supercritical water gasification process requires detailed knowledge of chemical reaction pathways, rates, and product yields. Laboratory-scale reactors are employed to develop this knowledge base. The rationale behind designs and component selection of continuous flow, laboratory-scale supercritical water gasification reactors is analyzed. Some design challenges have standard solutions, such as pressurization and preheating, but issues with solid precipitation and feedstock pretreatment still present open questions. Strategies for reactant mixing must be evaluated on a system-by-system basis, depending on feedstock and experimental goals, as mixing can affect product yields, char formation, and reaction pathways. *In-situ* Raman spectroscopic monitoring of reaction chemistry promises to further fundamental knowledge of gasification and decrease experimentation time. High-temperature, high-pressure spectroscopy in supercritical water conditions is performed, however, long-term operation flow cell operation is challenging. Comparison of Raman spectra for decomposition of formic acid in the supercritical region and cold section of the reactor demonstrates the difficulty in performing quantitative spectroscopy in the hot zone. Future designs and optimization of continuous supercritical water gasification reactors should consider well-established solutions for pressurization, heating, and process monitoring, and effective strategies for mixing and solids handling for long-term reactor operation and data collection.

## 1. Introduction

Supercritical water exhibits unique, tunable physiochemical properties beneficial for waste treatment, organic compound gasification, and material synthesis. At temperatures and pressures above the critical point (374°C, 22.1 MPa), properties can be varied continuously from liquid-like to gas-like without a phase change. A significant decrease in dielectric constant and quantity of hydrogen bonds across the critical point causes water to transition from a polar to a non-polar solvent [1]. Industrial applications of supercritical water require an understanding of physical phenomena linked to pressure, temperature, transport properties, compound solubility, and chemical reactions. Reactor optimization involves the selection of components, material, geometry, mixing strategy, instrumentation, reactor control, and catalyst introduction for an efficient and effective gasification process [1].

[1] Corresponding author, ivn@uw.edu

Supercritical water was first explored as a useful reaction medium in the 1970s for the hydrothermal refining of organic compounds to gaseous products [2]. Laboratory-scale supercritical water reactors (SCWRs) have been used to study supercritical water oxidation (SCWO) and gasification (SCWG) of model compounds, biomass feedstocks, and chemical warfare agent (CWA) surrogates, and have led to the development of pilot and industrial scale reactors [3]. SCWRs have also been constructed for hydrothermal synthesis of metal oxides and metal-organic frameworks, although such particle formation is undesirable in SCWO and SCWG reactors [4, 5].

SCWO is primarily used for neutralizing toxic waste. Under supercritical conditions, organic compounds and oxygen become fully miscible in water, allowing oxidation to occur in a single fluid phase with excellent transport properties. Many organic compounds are completely oxidized in under one minute with optimized temperature. Processing wastewater and sewage has been a common application for SCWO [6].

SCWG facilitates the decomposition of organic molecules in the absence of excess oxygen, through reductive hydrothermal reactions. Figure 1 displays a generic reaction network for the gasification of complex organic molecules. Desirable products from the SCWG of organic feedstocks include light fuel gases, such as $H_2$, CO, and $CH_4$; these pathways are favored at high temperatures and low feedstock concentrations. Common refractory products include phenolic compounds, furfurals, and char, which manifest due to polymerization reactions or the inability to break aromatic rings [7]. Metal oxides and inorganic acids can also form during SCWG, if heteroatoms or metals are present in the feedstock. Most SCWG research is focused on processing biomass [3, 8, 9], sewage sludge [10-12], or other organic wastes for conversion to drop-in fuels. An effectively designed SCWG reactor may be able to produce $H_2$ and $CH_4$ from waste biomass with >20% solid content at an economically competitive cost [11]. Biomass gasification in supercritical water is a carbon neutral process and requires no environmentally hazardous chemicals [5]. One major advantage of SCWG is that wet biomass can be readily gasified without an energy intensive drying step. The essential elements of a continuous SCWG reactor are presented with a representative schematic in Figure 2. Water and reagent are pressurized and subsequently heated to supercritical conditions, after which the reactor effluent is quenched and throttled back to ambient

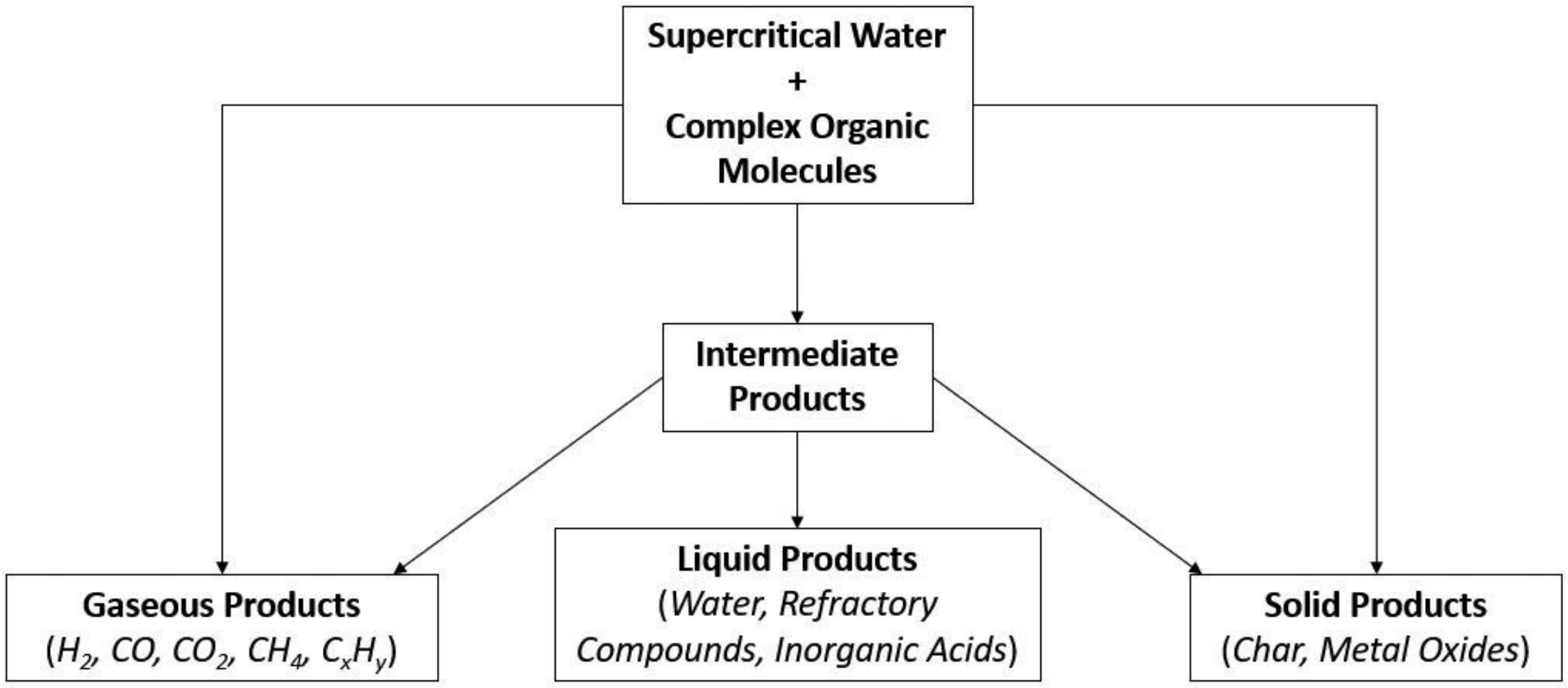

**Figure 1**: Generalized reaction network for supercritical water gasification of complex organic molecules. Desirable reaction pathways lead to high yields of gaseous products, while undesirable polymerization reactions lead to the formation of char. Heteroatoms or metals present in the feedstock can form metal oxides or inorganic acids, causing significant issues with reactor clogging and corrosion

conditions. While design choices vary considerably between individual reactors, the sequence of pressurizing, heating, reacting, quenching, and throttling remains consistent across all continuous SCWG reactors.

Despite similarities between SCWO and SCWG reactors, the design challenges are substantially different. The SCWO environment is oxidative, while the SCWG environment is reductive, thus the corrosion behavior of many materials differs between SCWO and SCWG [13, 14]. Some reactors have been designed to operate with or without an oxidant, such as the Sandia Supercritical Fluids Reactor (SFR) [15], but most are tailored towards a single processing regime.

SCWO is more promising commercially for hazardous waste treatment, as the exothermic oxidation process reduces fuel value in favor of high reaction temperatures and efficient chemical destruction [16]. A properly functioning SCWO system may have destruction and removal efficiencies (DRE) in excess of 99.999% and can readily handle most organic waste streams [14]. SCWO is appropriate when highly effective waste treatment, not fuel generation, is the primary goal. By contrast, SCWG allows for recovering much of the fuel value in the waste stream. SCWG is less effective at handling complex waste streams, due to issues with char formation and salt precipitation, but does promise economic benefits in fuel reforming. For industrial systems, heat recovery is one clear example where the two technologies require different approaches. Oxidation is an exothermic reaction scheme, while gasification facilitates endothermic reactions; heat recovery in a SCWG system is vital to process economics [3, 13]. A partial oxidation regime could improve gasification efficiency by facilitating internal reactor heating while some fuel value of the waste can still be recovered, and could offer an ideal middle ground between the two technologies [7, 17]. Advanced knowledge of heating values and chemical kinetic rates would significantly improve process designs. Only design strategies for SCWG reactors are considered in this paper. Some recommendations may be applicable to SCWO, but specific challenges will not be addressed, such as methods for oxidant introduction and thermal management of the reactor. Table 1 presents common design challenges for SCWRs, and compares how the challenges are addressed in SCWG and SCWO reactors.

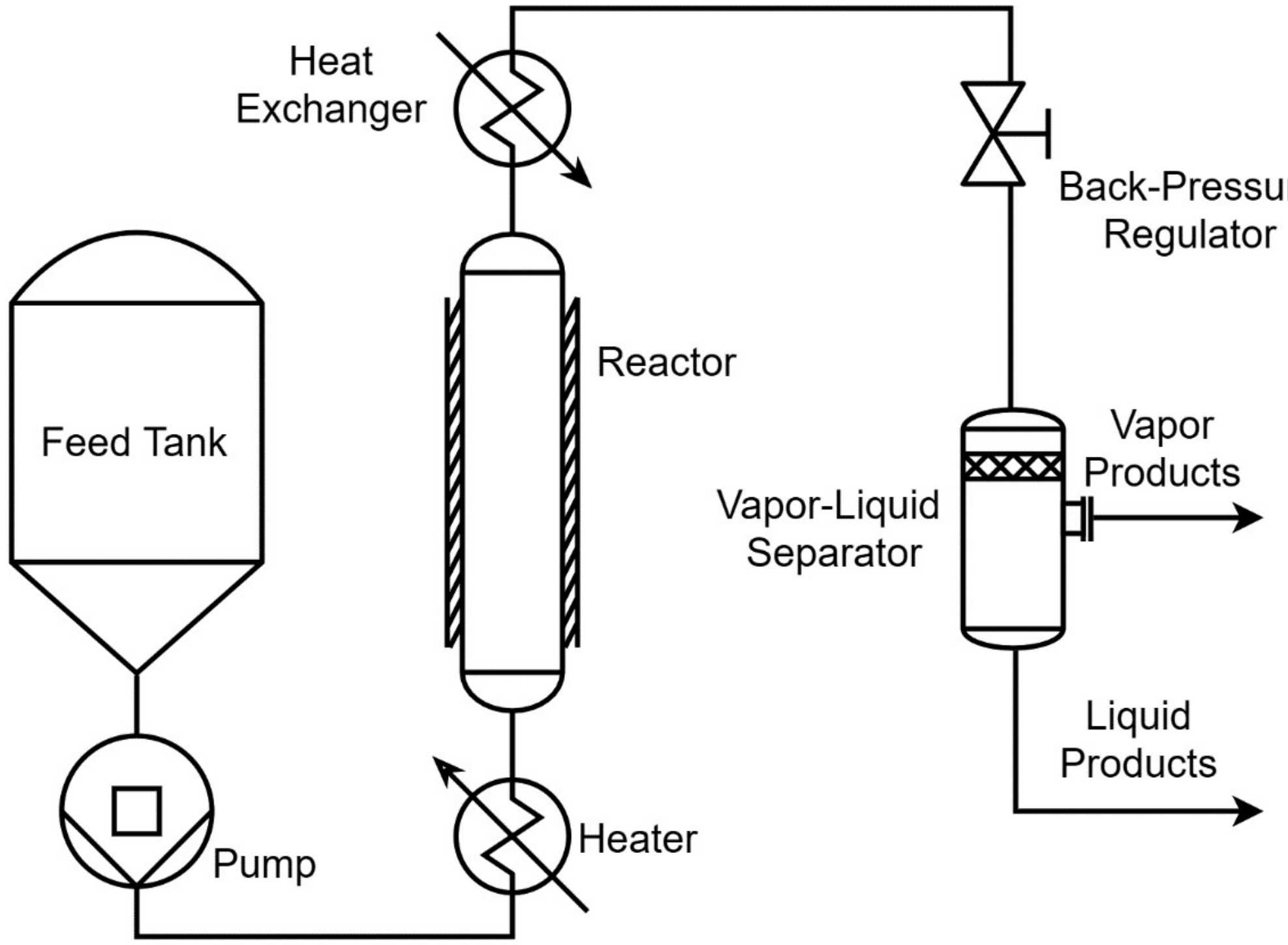

**Figure 2**: Representative schematic of a continuous supercritical water gasification reactor with premixed water and reagent

For SCWG, the yield of product gases is closely linked to reaction rates along competing pathways. Maximizing $H_2$ production is a common goal for biomass gasification; insight into reaction pathways and rates allows for the optimization of reaction temperatures, residence times, and selected catalyst [6, 18, 19]. A recent review of reported gasification rates and pathways in supercritical water has summarized how product composition and yields depend on reaction temperature, residence time, and feedstock concentration [20].

Although not considered in this paper, batch reactors have also been employed for lab-scale studies. Batch experiments require a fixed quantity of water and reagents to be pressurized, heated, and allowed to react at fixed conditions for a set amount of time before product collection and analysis. Batch reactors have different catalytic behavior than continuous flow reactors but offer insight into pure gasification routes. Continuous reactors are predominantly constructed from nickel-base alloys, which are known to catalyze gasification reactions. Catalytic behavior in batch reactors is either absent (i.e. quartz reactor) or limited by mass transfer rates due to the absence of flow in a batch regime [21]. Chemical kinetic rates determined in batch reactors are vastly different than rates determined in continuous flow reactors, and designs of each are markedly separate.

| **Challenge** | **Supercritical Water Gasification** | **Supercritical Water Oxidation** |
|---|---|---|
| Destruction of Refractory Compounds | High reaction temperatures and long residence times needed, high potential to form char [7] | Oxygen free-radicals facilitate efficient destruction, less potential to polymerize molecules and form char [16] |
| Fuel Value Recovery | Reductive reactions allow for recovery of feedstock fuel value in gaseous form [1, 3, 6] | Oxidative reactions consume feedstock fuel value in favor of compound destruction [14, 16] |
| Reactor Thermal Management | Endothermic reactions necessitate additional heater(s) to maintain isothermal conditions | Cooling system or sand bath needed to prevent thermal runaway during exothermic reactions |
| Corrosion | Heteroatoms and salts are highly corrosive [13, 14] | Heteroatoms and salts are highly corrosive, oxide layer forms on metal reactor walls [13, 14] |
| Clogging | Char formation likely from complex organic feedstocks, salt precipitation and metal oxide formation commonly causes clogging [1, 7] | Salt precipitation and metal oxide formation commonly causes clogging [16] |
| Process Economics | >20% solid content and efficient heat recovery needed for cost-effective fuel gas production [3, 11, 13] | Regenerative heating minimizes need for external energy input |
| Practical Application | Fuel gas production from wet organic wastes (e.g. sewage, biomass) [3, 8-12] | Destruction and removal of toxic compounds (e.g. sewage, CWAs) [14] |

**Table 1:** Comparison between operational challenges in continuous SCWG and SCWO reactors

This review provides thorough insight into known design solutions for lab-scale, continuous SCWG reactors, which allows researchers to approach the design of new reactors with a clear understanding of foundational knowledge. Discussion on reactor subsystems provide suggestions for specific components and suppliers for effective pumps, heaters, and back pressure regulators. Common design challenges are discussed, including (i) corrosion of reactor components, (ii) reaching and controlling supercritical pressures and temperatures within the reactor, (iii) clogging due to char formation or salt precipitation, and (iv) acquiring time-resolved chemical species information for process control or reaction rate analysis. In this manuscript, we provide insight into potential solutions for each of these challenges. All told, a state-of-the-art perspective on common designs, challenges, and uses of lab-scale, continuous SCWG reactors is presented.

# 2. Main Text

Costs and research goals often influence design choices for lab-scale reactors. Corrosion mitigation, pressurization, heating methods, effluent quenching, data acquisition, and mixing must be considered in all designs; some generalized solutions have been identified in published scientific reports. Open questions still remain in the best methods to pretreat solid feedstocks, and how to properly mitigate clogging due to solid formation within the reactor.

## 2.1 Feed Pretreatment

Effective pretreatment of real gasification feedstocks is necessary to facilitate SCWG at the industrial scale. Pretreatment is rarely necessary for model compound gasification, aside from the dissolution of the compound into distilled, deionized water. For solid model compounds, gasification is performed at feedstock concentrations at or below the saturation limit of the compound in room temperature water. For liquid compounds the feedstock concentration can be as high as experimentally desired.

Insoluble solid or viscous feedstocks require pretreatment to facilitate gasification. For biomass feedstocks, small particles are needed to create a pumpable slurry. These can be obtained by crushing or milling the biomass [1, 3]. Subsequently, the biomass is sieved or filtered to remove large particles [22]. Water content can be adjusted to make the pressurization step easier and to reduce the potential for char formation [3, 22]. Pump selection and dry matter content of the biomass slurry must be considered. Faires [23] describes an approach where two cylinder piston pumps with a Y-coupler continuously pump a biomass slurry with 15% solid content to 27 MPa at a flow rate of 5 g/s. The maximum particle size was 0.84 mm, as the ground biomass was sieved with mesh No. 20 before mixing with water.

Matsumura [24] has successfully used a liquefaction pretreatment to facilitate pumping complex biomass, by heating the reactant stream to 200°C prior to pressurization. This hydrothermal pretreatment softens the biomass by breaking down the cellulosic structure [25]. Alternatively, researchers at the University of Hawaii have reported successful gasification of sawdust by suspending the sawdust in a dilute starch gel before pressurization [25]. Finally, flash pyrolysis has been used by Penninger et al. [26, 27] as an effective biomass pretreatment. Flash pyrolysis involves rapid heating of the feedstock at 500°C to 600°C for a few seconds, which allows for capturing sand and minerals in the produced char and ash. Additionally, the resulting carbonaceous products are liquid and easily pressurized.

## 2.2 Corrosion Mitigation and Material Selection

Corrosion control methods in SCWRs have been extensively studied [13, 28-35]. In general, the high pressure, high temperature (HTHP) supercritical water environment is corrosive to most materials, particularly if heteroatoms or alkali salts are present [36]. The heat exchanger is the system component most subject to corrosion, due to the transition through the highly corrosive critical region. Chlorinated compounds are especially corrosive, although they are typically absent from standard SCWG feedstocks [13]. Four primary categories of corrosion mitigation methods have been described in a review from Marrone et al. [13]. These include (i) preventing corrosive species from interacting with the reactor surface, (ii) forming a corrosion-resistant barrier, (iii) selecting materials resistant to corrosion, and (iv) tuning operating conditions to avoid severe corrosion conditions. These approaches can be combined within the same system. While useful in large-scale SCWRs, some of these techniques are not appropriate for lab-scale reactors. For example, in a reactor designed for the study of decomposition rates, preventing contact with reactor walls would introduce a concentration gradient within the flow, invalidating the assumption of a plug-flow regime. Additionally, forming a corrosion resistant barrier on the interior of the reactor wall would reduce the catalytic wall effect. For lab-scale reactors, the most practical corrosion mitigation technique is appropriate material selection.

Certain nickel-base alloys resist corrosion in the presence of salt precipitates that are likely to form during the gasification of complex feedstocks. The majority of continuous SCWRs have been constructed from Hastelloy C-276 or Inconel 625 due to the combination of corrosion resistance, strength at high temperatures, and commercial availability. Inconel or Hastelloy tubing and fittings can be purchased from such companies as High Pressure Equipment Co. (Erie, PA) and Swagelok (Solon, OH). Tang et al. [14] demonstrated that these alloys tend to gain mass in the presence of corrosive species, while stainless steel alloys lose mass. A buildup of deposits can be cleaned periodically, but lost mass would eventually lead to system failure. Titanium alloys were shown to resist corrosion under supercritical conditions, but titanium does not provide beneficial catalytic activity [14].

Another solution to mitigate corrosion is to manufacture a removable reactor liner, which can be cleaned or replaced periodically. Titanium liners were successfully used in SCWO reactors designed for CWA destruction, and this solution could be applied to SCWG reactors [37, 38]. However, a liner would provide additional resistance to heat transfer into the gasification environment [13]. The introduction of a catalyst may reduce corrosion by lowering operating temperatures, but process complexity is increased.

## 2.3 Pressurization

Any pressurization method for SCWRs should be capable of reaching pressures above the critical point, and allow for a range of user-specified flow rates. Among reviewed systems, the single method used for independent mass flow and pressure control is to use constant flow rate pumps in conjunction with a back-pressure regulator (BPR). Positive displacement pumps are effective at the laboratory scale for consistent flow rates of simple feedstocks at high pressure.

Due to their compact size, commercial availability, integrated controls, and excellent reliability, high-performance liquid chromatography (HPLC) pumps are often used when pumping simple liquid feedstocks at the lab-scale [18, 39-48]. Off-the-shelf HPLC pumps are available from Waters [46-48] (Milford, MA), Eldex [18] (Napa, CA), Knauer [39] (Berlin, Germany), Teledyne SSI [42] (State College, PA), and JASCO [43] (Easton, MD), among other companies. HPLC pumps offer precise control over flow rates typically between 0.01 to 10 mL/min. HPLC pumps are not suited for large reactors and are not effective for pumping slurries [46].

Diaphragm pumps are ideal for pressurizing slurries with high solid content or highly viscous feedstocks. However, diaphragm pumps do not offer as precise of flow rate control as HPLC pumps. Ondze et al. [23] report using a diaphragm pump from LEWA (Leonberg, Germany) in their experiments gasifying beet residues, and Caputo et al. [39] report using a Milton Roy (Houston, TX) high-pressure membrane pump to introduce a glucose solution into their gasification reactor. Klingler et al. [67] used diaphragm pumps from Orlita and LEWA. Xu et al. [46] relied on a diaphragm pump to pressurize sewage sludge for gasification, although an exact model was not specified.

Syringe and plunger pumps are sometimes used in lab-scale SCWG reactors [42, 49, 50]. While less precise than HPLC pumps, syringe and plunger pumps offer higher flow rates, although they cannot handle slurries with high solid content or large particle diameters [42, 50]. Molino et al. [50] report using a Teledyne (Thousand Oaks, CA) syringe pump with flow rates up to 204 mL/min. Elliott et al. [49] detail using a reciprocating plunger pump in their SCWG reactors, although flow rate capabilities, make, and model are not specified.

## 2.4 Heating

At 25 MPa, increasing the temperature of water from 20°C to 400°C requires a heat input of 2483 kJ/kg [51, 52]. This can be accomplished in different ways, with three common strategies identified in SCWG systems: immersive baths, resistive contact heaters, and radiative furnaces.

Supercritical water flowing through heated tubes can exhibit two irregular heat transfer regimes. Enhanced heat transfer (HTE) and deteriorated heat transfer (HTD) are characterized by exceptionally efficient or poor heat transfer to the flow, respectively. Near the critical point, HTE occurs due to an optimal combination of thermophysical properties such as density and specific heat. Alternatively, HTD occurs due to a sub-optimal combination of fluid properties. HTD can cause wall temperatures to increase well above the fluid temperature, potentially damaging the reactor walls. Smaller laboratory scale reactors operating at lower flow rates may suffer from HTD; the reduced heat-transfer coefficient at low mass fluxes appears to arise due to buoyancy effects and flow acceleration due to property changes in the heating section. Potential solutions for avoiding HTD include orienting the flow downward to take advantage of buoyancy and optimizing mass flux to heat flux ratio through selection of tubing diameter [53].

Electric furnaces can be purchased with high power ratings and built-in controls, making them a convenient heating solution [15, 41, 42, 48-50, 54-56]. Some furnaces contain multiple, independent heating zones for precise temperature control. Many come fully insulated. Due to the high operating temperatures of the heating elements, electric furnaces are not likely to overheat. Care must be taken not to exceed the safe operating temperatures of the reactor material, as many furnaces can heat the reactor walls to unsafe temperatures. Some companies offering radiative tubular furnaces with power ratings suitable for lab-scale SCWRs include Thermcraft [15, 41] (Winston Salem, NC), Applied Test Systems [42] (Butler, PA), Vecstar [50] (Chesterfield, UK), and OMEGA [56] (Stamford, CT).

Contact electric resistive heaters transmit heat through conduction. Various form factors exist, including band heaters, cartridge heaters, heating coils, and flexible heating cables. Contact heaters require external insulation to minimize heat loss but tend to be cheaper and smaller than electric furnaces or fluidized baths. Resistive heaters often contain embedded thermocouples for feedback temperature control. As long as the temperature in the heating element is kept below the temperature limit of the wall material, the system can operate safely. To make up for the low maximum temperature limit of many resistive heaters, a second stage heater frequently provides additional energy [15, 39, 48, 54, 56, 57]. The most significant drawback of this type of heater is an inability to operate at high temperatures in sections of the reactor experiencing HTD. Typical solutions require either a longer heating section or the use of an alternative heating method for the supercritical region. Resistive heaters are available from WATLOW [15] (St. Louis, MO), OMEGA, and other companies.

A unique heating approach was used in heat transfer experiments in a vertical bare tube reactor at the State Scientific Centre of the Russian Federation (Obninsk, Russia). The reactant was heated by passing an electrical current through reactor walls [58, 59]. While efficient, the 4-meter vertical tube reactor consumed large amounts of electrical power.

Immersion heating refers to heating the reactor section in a fluidized bath, ensuring isothermal conditions and good heat transfer. Size and cost have prevented immersion heaters from being widely adopted for lab-scale reactors. Some designs utilize a hybrid approach by bringing the water to the supercritical state with a furnace or contact heater in the preheating section and using immersion heating for the reactor section. Immersion heating can be implemented with a fluidized sand bath [43, 60, 61], a molten salt bath [57], or another medium such as a fluidized alumina bath from Techne [40] (Staffordshire, UK).

## 2.5 Reagent Mixing

Two mixing strategies have been reported: (i) mixing water and reagents before heating, and (ii) injecting cold reagents into supercritical water. Post-critical injection can rapidly heat and mix reagents, creating a definitive reaction start time for chemical kinetic experiments. The design of the mixing section and operating conditions need to be considered to optimize the mixing rate. Numerical simulations can assist in the design of an effective mixer, but ambiguity in supercritical fluid properties and diffusion coefficients need to be considered [62].

Premixing allows for solid or viscous feedstocks to reach operating pressures as an emulsion. Water is necessary as a transport media for gasification when feedstocks are solid, viscous, or insoluble in room temperature water. One drawback to premixing is char and tar formation in the preheater, due to slow heating of the reactant [13, 39]. Char formation is reduced by rapidly heating the reagents to supercritical temperatures. A thorough understanding of char formation pathways and rates is needed to optimize heating of premixed slurries.

Post-critical injection of reagents significantly reduces char formation and allows for reasonably accurate calculations of residence time for chemical kinetic studies [8, 63, 64]. The design of an appropriate post-critical mixing section introduces new challenges. In systems with low flow rates, the high kinematic viscosity of supercritical water yields low Reynolds numbers, resulting in laminar mixing limited by molecular diffusion rates. In early research, the SCWR at MIT yielded inconsistent results due to inefficient mixing of water and reagents. When comparing SCWO rates of methanol with other research groups, the assumption of fast-mixing was found to be inaccurate. The Reynolds number in the original MIT reactor was around 3100; too low for consistent turbulent mixing [65]. A smaller diameter (0.25 mm) injector reduced the mixing time [66]. Larger reactors and higher flow rates promote mixing through an increase in the Reynolds number. In general, optimization of the operating conditions and the critical reactor dimension is required to achieve adequate mixing profiles for plug flow conditions in the reactor section. Most authors do not report mixing section design, but variation in mixing profiles could explain discrepancies in reported reaction pathways and rates [20].

The two mixing scenarios have an analogy in combustion research where two types of mixing are often used: (i) premixed flames and (ii) diffusion flames. The overall rate is often computed based on the limiting rate approach, and careful evaluation is needed to ensure that the reaction rate, not the mixing rate, is the limiting rate during experiments. Numerical and experimental studies could lead to understanding of the optimal mixing section design, which could be standardized in future studies [39]. The mixing limit (both thermal and species concentration) is specifically important for high feed concentrations and regions with competing reaction pathways and/or fast kinetic rates. The effect of the mixing rate on the overall reaction can be significant for slow mixing and fast chemical rates. In laminar flows, the mixing is driven by molecular diffusion and can be calculated from first principles. Since very few experimental datasets are available [67], analytical modeling can be used for some properties based on information from the NIST database [68]. In turbulent flow, molecular diffusion is relevant in the viscous sublayer, especially when catalytic reactions may occur at the reactor wall.

The relationship between the mixing and kinetic rates can be described by the global and local Damkohler (*Da*) and Karlovitz (*Ka*) numbers, which are based on the ratios of chemical and mixing timescales and are often used in combustion modeling. The local Da is especially of interest as it shows the regions where the reaction is dominated by mixing or kinetic rates [69]. In numerical simulations of turbulent flows, the competition between the mixing and chemical kinetic rates in the presence of turbulence can be evaluated using a model describing turbulence-chemistry interaction, such as the eddy-break-up model or the eddy dissipation concept presented in combustion literature [70-72]. The mixing rate limit is specifically important in the design of a practical SCWG system. The application of limiting rate models to the laboratory and industrial reacting systems can be found in [73-75].

## 2.6 Clogging and Salt Precipitation

A significant challenge with SCWG of real feedstocks is reactor clogging, either from char buildup or precipitation of insoluble compounds. This issue is especially pronounced in reactors with a packed catalyst bed, where the free flow diameter is significantly reduced [1]. Salts, either added as catalysts or naturally present in biomass feedstocks, have low solubility in supercritical water [1]. Alkali metal salts exist in a molten form at temperatures above 300°C, and tend to adhere to reactor surfaces, corroding the metal and poisoning catalytic pores. Potassium is often present in biomass feedstocks, thus potassium carbonate ($K_2CO_3$) and potassium bicarbonate ($KHCO_3$) have been used as model salts in gasification studies of model biomass compounds. Buildup can be cleaned by periodic flushing with cold water, but frequent cooldown of the system is impractical. Salt precipitation can be suppressed by avoiding the addition of alkali metal salts, or by pretreatment to remove alkali compounds.

Many biomass constituents react via two competing reaction pathways, one leading to gaseous products and one to char [7]. The gas formation pathway is favored by fast mixing and heating to supercritical temperatures, by the presence of effective catalysts, and by operating at temperatures well above the critical point [7, 20]. Low feedstock concentrations suppress char formation, although reducing feedstock concentrations is not practical for large-scale systems [20]. Char can be gasified slowly, although the ideal solution is to avoid its formation altogether.

Researchers at the University of Texas at Austin investigated methods for separating solid particles within the gasification environment in cases where solid formation is inevitable. Orienting the reactor downward can help manage solid precipitation, however more involved separation methods may be needed for some feedstocks [7]. In 1993, a hydrocyclone was tested for particle separation in the supercritical environment, with separation efficiencies ranging from 80% to 99% depending on the tested particulate [76]. Crossflow microfiltration was also tested for the removal of inorganic salts and metal oxide particles. The crossflow microfilter was found to be 40% to 85% efficient at separating inorganic salts, depending on the reactor temperature [77]. Separation efficiencies were well in excess of 99% for the removal of metal oxide particles [78]. Any separation method in the supercritical environment must be corrosion resistant and must minimize interaction of molten salts with reactor components.

## 2.7 Heat Exchanger and Back Pressure Regulator

Once the supercritical effluent exits the reactor section, it is typically quenched by a heat exchanger and throttled to atmospheric pressure by a back pressure regulator (BPR). The design of the heat exchanger is subtly important for chemical reaction studies. *In-situ* Raman spectroscopy has demonstrated that for a heat exchanger that is not oriented vertically, flow separation can lead to accumulation of insoluble gases within the heat exchanger. Separation occurs when the effluent transitions to a two-phase flow below the critical point. At subcritical temperatures in the range of 240°C to 260°C, formic acid was shown to be an intermediate of the water-gas shift (WGS) reaction [79]. The synthesis of formic acid in the heat exchanger would lower $H_2$ yields and would lead to inaccuracies in chemical kinetic studies. The heat exchanger should be oriented vertically, taking advantage of buoyancy effects to allow insoluble gases to exit [56].

For large scale systems, heat recovery is vital to process economics. For a feedstock with water content above 80%, often the energy content of the feedstock is lower than the energy required for the water to reach reaction conditions [3]. However, at the lab-scale, most groups separate the heating and cooling systems, to simplify component selection.

BPR selection is important, as solid precipitates could clog the ports. The wetted material on the BPR must be compatible with corrosive reaction products or refractory organic solvents in the effluent stream. Diaphragm style BPRs are the most common choice for SCWG reactors, with pressure control from spring loading or from dome loading. Dome loading offers precise control via an external pressure source, such as compressed $N_2$. Dome loaded BPRs are available from companies such as Equilibar [56] (Fletcher, NC). However, the cost tends to be higher, and the reactor form factor increases due to the external gas tank. Spring loaded BPRs offer less precise control but are less expensive and more compact. Spring loaded BPRs are available commercially from TESCOM [42] (St. Louis, MO), Swagelok [57] (Solon, OH), and other companies. Studies have shown that pressure plays a negligible role in gasification rates, thus spring loaded BPRs are acceptable for most SCWG reactors [7, 57]. High-precision pressure control is only necessary for studies performed near the critical point, where small pressure fluctuations can impact the density of the supercritical water [80].

## 2.8 Reactor Monitoring, Control, and Data Acquisition

Real-time process monitoring allows for the safe operation of SCWG systems that are prone to clogging, corrosion, leaks, or structural failure. Knowledge of temperature, pressure, residence time and chemical composition are required for chemical reaction rate studies. Some standard approaches are reported in the literature. Thermocouples immersed in the flow directly measure the reactor temperature with a fast response time but are exposed to corrosion. Thermocouples affixed or embedded external to the reactor avoid corrosion but respond slowly and do not directly measure fluid temperature. External thermocouples can be used to monitor the temperature of heating elements and reactor walls to ensure that safe operating temperatures are not exceeded. A combination of internal and external thermocouples allows for thorough monitoring of reaction conditions and component temperatures.

Commercially available pressure sensors are not suited for supercritical water environments. Reactor pressure must be monitored in the cold zones, preferably before the preheaters, where a sudden rise in pressure would indicate a clog downstream. The installation of rupture discs or pressure relief valves on the front end of the reactor is recommended to avoid system failure in the case of over-pressurization.

### 2.8.1 Residence Time

To estimate the reaction residence time, researchers typically assume a plug flow regime in the reactor, removing the need to consider concentration gradients, fluid property variations, or residence time variations. Molecular concentrations in the direction of the flow are assumed to be a function only of residence time. In the absence of turbulence, species transport is governed by molecular diffusion, which does not facilitate rapid mixing of reagents. Tiwari et al. described the mixing behavior of benzene in supercritical water under laminar conditions [62]. Computational fluid dynamics (CFD) were used to compare the simulation results for the binary diffusion coefficient from experimental studies and kinetic theory parameters presented in literature [67, 81]. The study demonstrated that variations in residence time can exist in the reactor section with the fluid near the wall having greater residence times than the fluid in the center of the tube (3.05 mm ID tube was modeled).

One strategy to obtain uniform residence time and species concentrations is to induce secondary Dean flow by coiling the reactor. For Dean numbers (De) > 75, the secondary cross-sectional motion is superposed on the primary flow as a pair of counter-rotating cells [82, 83]. Dean vortices in the cross-sectional direction have been shown to enhance mixing [84-86], and more recently this approach was applied to microfluidics where mixing is particularly challenging due to the very low Reynolds number [87, 88]. Interestingly, the pipe curvature maintains laminar flow at greater Reynolds numbers than for straight pipes, despite curvature being known to cause instability [89]. Optimization of operating parameters is necessary to estimate mixing levels in the reactor. The use of an analytical or numerical model is required to address these issues.

### 2.8.2 Ex-situ product collection and analysis

The vast majority of SCWG studies rely on *ex-situ* analysis to identify reaction products and compound yields. *Ex-situ* analysis may simply involve quantifying gaseous product yields, or it may involve identification and quantification of every product in the effluent. Researchers attempting to determine underlying chemical mechanisms and kinetic rates must identify all reaction products for each experimental condition. Typically, a gas-liquid separator exists after the back pressure regulator. Gas products are often analyzed using gas chromatography (GC) with a thermal conductivity detector (TCD) and flame ionization detector (FID), to quantify yields of $H_2$, CO, $CO_2$, $CH_4$, and occasionally trace amounts of $C_2H_4$ and $C_2H6$ [9, 11, 12, 15, 18, 19, 22, 26, 27, 39, 41-43, 46-48, 50, 60, 61, 98-104, 106].

GC is not sufficient for understanding chemical kinetics, as liquid product yields must also be considered. A total organic carbon (TOC) analyzer allows for calculating carbon conversion efficiency by quantifying the concentration of carbonaceous compounds in the liquid effluent [9, 12, 15, 18, 22, 39, 40, 44-48, 54, 63, 98-100, 103, 104, 107]. Identification and quantification of liquid products with HPLC, NMR spectroscopy, Fourier transform infrared (FTIR) spectroscopy or Raman spectroscopy allows for identifying and quantifying liquid product yields, which can be used to understand chemical reaction pathways and rates [12, 19, 40-42, 44-46, 50, 54, 57, 63, 102, 104]. Occasionally solid products are analyzed with such measurement techniques as scanning electron microscopy (SEM), proton-induced X-ray emission (PIXE), FTIR spectroscopy, and other methods [9, 11].

While straightforward, *ex-situ* analysis is time consuming, and not ideal for chemical kinetic studies. *Ex-situ* analysis is most suited to studies where only gasification efficiency is measured, and a deeper understanding of reactions mechanisms is not sought.

### 2.8.3 In-situ monitoring

*In-situ* analysis allows for analysis of reaction products in real time. Raman spectroscopy with an excitation wavelength in the visible range is one of the most promising *in-situ* process monitoring techniques for effluent composition analysis due to the large spectral window in supercritical water. Raman spectroscopy in a continuous flow SCWR was first demonstrated by Sandia National Laboratories to investigate the oxidation kinetics of methanol and isopropyl alcohol [15, 90, 91]. Data collection within the reactor allowed for experimental residence times as low as 0.1 s. However, long-term use of the optical cell in the HTHP environment led to cell failure due to thermal expansion and thermal cycling. *In-situ* Raman cells require optical access via one or more window. The optical cell in the SFR consisted of a sapphire plug sealed using a gold gasket and a backing nut, but the gasket would creep during extended use, or the sapphire window would break under thermal stresses. Bellville washers were added to relieve thermal stress, but creep remained an issue. Use of the Raman cell in the HTHP environment was ultimately abandoned [92].

The University of Tokyo utilized *in-situ* Raman to study the molecular structure of methanol [93] and oxygen [94] in supercritical water, as well as to measure the kinetic rate of methanol oxidation [95]. The authors did not report any difficulties with maintaining a seal in the optical cell, and details on the exact design of the cell are sparse. From figures provided, it appears that the Raman cell uses a cylindrical sapphire window, held in place by a gasket and a backing nut, which is very similar to the design used by Sandia.

The prospects for Raman spectroscopy in HTHP environments were summarized in a recent review [96]. Raman is acknowledged to be a promising technology for revealing qualitatively and quantitatively how biomass decomposes in supercritical water, and how and when species form from intermediate compounds. Raman is also identified as a possible tool for real-time pocess control, as it could be integrated in a feedback control loop with control decisions based on product mixture composition. The author acknowledged that optical access is a significant challenge, and suggested the use of micro capillaries made of glass for optical access to the HTHP environment.

The University of Washington SCWG reactor uses an immersion Raman probe, where a sapphire ball lens provides optical access and laser focusing. The sapphire ball is fixed in place by a backing nut and two gold gaskets. High-resolution Raman spectra can be obtained in the backscatter configuration, and this technology has been used to study formic acid gasification in near-critical and supercritical water.

Placing the Raman probe in the hot zone of the reactor allows for precise residence time calculations, as data is collected at a defined location. Residual reactions occurring within the heat exchanger do not affect collected data. Additionally, Raman monitoring in the hot section theoretically allows for identification of unstable intermediate reaction compounds. However, no reliable HTHP optical cell is commercially available. A suitable cell must be custom-built. Additionally, for quantitative spectroscopy, concentration measurements are determined with models that must first be calibrated with known mixture spectra. Due to non-linear spectral effects and variations in fluid density at supercritical temperatures, calibration spectra would need to be collected at supercritical temperatures and pressures. For reactive organic species, these calibration spectra would be practically difficult to generate.

Raman spectroscopy in the cold zone allows for data collection at a consistent temperature and pressure. *In-situ* monitoring after the heat exchanger is sufficient for determining global decomposition rates, as hydrolysis, decarboxylation, and dehydration reactions are quenched at low temperatures. Calibration spectra must only be collected at high pressure. Thermal stresses are avoided. However, residence time calculations are less precise due to the time necessary to quench the hot effluent. Low temperature reactions will continue through the heat exchanger, such as water-gas shift and methanation reactions, slightly affecting the observed product concentrations. Finally, flow separation in the heat exchanger, as mentioned previously, can complicate concentration measurements of insoluble product gases.

Formic acid was gasified in the University of Washington SCWG reactor at known temperatures and flow rates, to determine the significance of placing the immersion Raman probe in the hot zone of the reactor, compared to the cold zone after the heat exchanger. Figure 3 shows the baseline-corrected Raman spectra collected in the cold zone, and in the hot zone at operating temperatures of 365°C, 380°C, and 400°C. Data collection in the hot zone did not identify additional intermediate reaction products. At 365°C reaction products are clearly visible, although peaks have shifted relative to known positions at cold temperatures. At temperatures of 380°C and 400°C, the signal of reaction products is significantly diminished. This is hypothesized to be caused by a loss of optical access to the reaction zone. Sapphire is visible, proving that the optical path is still intact. Likely, thermal effects are altering the spectral transmission of the Raman laser through the ball lens. Hanush et al. [15] note that realignment of the Raman laser was frequently needed when operating the Sandia reactor at supercritical conditions, either due to laser misalignment or loss of transmission through the sapphire window, supporting the theory that loss of transmission is occurring through the sapphire ball lens. Regardless, collected spectra demonstrates that extracting quantitative data from HTHP Raman spectra is practically difficult, due to peak shifting and potential temperature-dependent optical effects. Any HTHP Raman system must allow for laser realignment in case of difficulty with optically accessing the reaction environment.

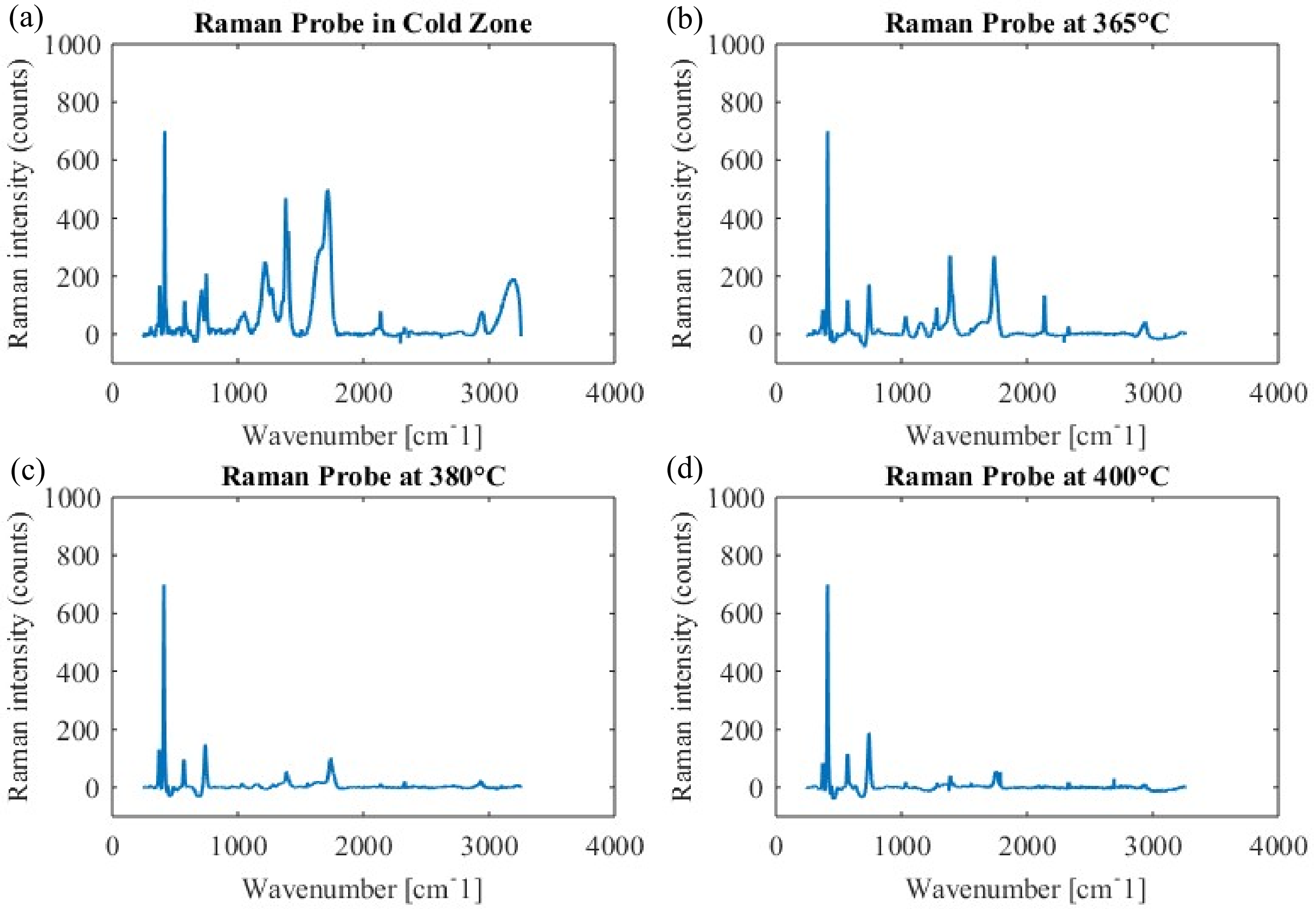

**Figure 3**: Plots of baseline-subtracted Raman spectra from formic acid decomposition experiments with (a) Raman probe in cold zone, (b) Raman probe in hot zone at 365°C, (c) Raman probe in hot zone at 380°C, and (d) Raman probe in hot zone at 400°C. Significant reduction in signal intensity appears to be due to temperature-dependent optical effects, which significantly complicate quantitative spectroscopy

## 2.9 Catalysts

Motivated by the high temperatures and low concentrations needed to fully gasify biomass in supercritical water, the introduction of homogenous or heterogenous catalysts has been investigated as a method to improve process economics. A review by Guo et al. [97] highlights the effectiveness of common catalysts. Most catalysts favor the WGS reaction, which improves $H_2$ yield. However, some catalysts do not speed underlying gasification reactions such as hydrolysis, decarboxylation, and dehydration. Economic and life-cycle assessments are needed to determine whether catalyst integration significantly improves process economics or system longevity.

Alkali metals (i.e. $Na_2CO_3$, $KHCO_3$, and NaOH) are effective catalysts, however these compounds cause issues with corrosion, reactor fouling, and clogging due to the insolubility of salts in supercritical water. The mechanistic understanding of the catalyst is limited, due to the interaction of the salts with the reactor walls. It is thought that the alkali salts may dissolve a protective metal oxide layer on the inside reactor surface and resulting corrosion products themselves may act as catalysts [7]. The introduction of alkali metals for catalysis is not recommended [97].

Activated carbon (AC) effectively catalyzes the WGS reaction and the gasification of simple hydrocarbons, such as glycerol and glucose [46]. Yet it was found to be ineffective in catalyzing the gasification of glycine [98]. AC itself decomposes in supercritical water and deactivates within two to four hours of reactor use; it is not suitable for long-term reactor operation [46]. AC is often used as a support for metal-based catalysts but would be a poor choice for a general gasification catalyst [97, 99].

Nickel is catalytic to gasification but tends to sinter and deactivate during long-term use [21, 47, 97, 99-101]. A layer of carbon or salt precipitates can adsorb on the nickel surface, poisoning the catalytic effect [97]. Present in common reactor materials, the catalytic effect of nickel is typically active in continuous flow SCWG reactors whether it is desired or not. Small diameter tubing used for lab-scale reactors leads to higher surface-to-volume ratios (S/V), increasing the significance of the catalytic wall effect. Inducing Dean vortices, maximizing S/V, or selecting materials high in nickel content can maximize the beneficial catalytic activity of nickel components.

Ruthenium is the most catalytically active gasification catalyst, shown to effectively catalyze the gasification of glucose, microalgae, and glycerol with high stability [100-105]. Ruthenium was demonstrated to be effective in preventing the formation of char and refractory intermediate compounds from the gasification of biomass constituents [104]. However, ruthenium is expensive, and sulfur is known to poison the ability to catalytically cleave C-C bonds [97, 101]. Rhodium, platinum, cobalt, molybdenum, iridium, and palladium based catalysts have all been investigated, but are less effective than nickel or ruthenium in improving gasification efficiency or $H_2$ yield [97, 101, 106]. Table 2 presents further details on promising noble metal catalysts for SCWG.

Metal-based catalysts are most often impregnated on a support compound and packed into the reactor as pellets, held in place by porous frits [43, 46, 47, 97, 99, 100, 102, 103]. Potential support compounds include activated carbon, zirconium dioxide ($ZrO_2$), titanium dioxide ($TiO_2$), ceria ($CeO_2$) and aluminum oxide ($\alpha$-$Al_2O_3$) [43, 47, 97, 99-103, 106]. During operation the catalyst is often depleted or deactivated and cannot be replenished without complete shutdown and disassembly of the reactor. Depletion of the catalyst leads to metal compounds in the effluent, which are often toxic and must be dealt with through separation and/or neutralization [97].

| Catalyst/Support | Synthesis Method | Properties & Performance | Source |
|---|---|---|---|
| Ru/γ-$Al_2O_3$ | Commercially obtained | Highest catalytic activity for gasification of alkylphenols; decreased activity after transition from γ- to α-phase alumina; high activity for C-C bond cleavage | [18, 100, 103, 104, 108, 109] |
| Ru/$TiO_2$ | Commercially obtained | Highest catalytic activity for gasification of lignin; high activity for C-C bond cleavage | [110-112] |
| $RuO_2$ | Commercially obtained | Conversion superior to catalysis by NiO, $MoO_3$, and $ZrO_2$ | [105] |
| Ru/C | Commercially obtained | High catalytic activity; decreased activity after repetitive use | [109, 110] |
| Ni/γ-$Al_2O_3$ | Incipient wetness impregnation; $Ni(NO_3)_2{\cdot}6H_2O$ precursor | Highest catalytic activity and $H_2$ selectivity of 17 supported transition metal catalysts tested for SCWG of glucose in [18] | [18, 100] |
| Ni/$SiO_2$ | Evaporative deposition; $Ni(NO_3)_2{\cdot}6H_2O$ precursor | High $H_2$ selectivity; high activity for C-C bond cleavage | [113] |
| Pt/$SiO_2$ | Ion exchange at pH = 11; $Pt(NH_4)_4(NO_3)_2$ precursor | High $H_2$ selectivity; moderate activity for C-C bond cleavage; low methanation rate | [113] |
| CuO | *In-situ* hydrothermal generation of nanoparticles; $Cu(CH_3COO)_2$ precursor | High S/V ratio; effective catalyst for methanol reforming; not effective for cleaving C-C bonds of larger molecules | [107] |

**Table 2**: Promising metal catalysts for supercritical water gasification of complex organic feedstocks

A novel solution to the issue of catalyst depletion and deactivation is to continuously synthesize catalytic nanoparticles *in-situ*. The supercritical water environment can facilitate rapid synthesis of metal oxide nanoparticles, such as copper oxide (CuO), iron oxide ($Fe_2O_3$), nickel oxide (NiO), and zirconium dioxide ($ZrO_2$). Typically, a water-soluble metal salt precursor is rapidly mixed with supercritical water to facilitate synthesis. By synthesizing metal oxide nanoparticles upstream of the reactant, catalytically active nanoparticles with high surface-to-volume ratios can be continuously replenished in the reactor section. Gadhe and Gupta [107] demonstrated the efficacy of this process by continuously generating copper oxide nanoparticles *in-situ* from a feed of cupric acetate solution. The nanoparticles were shown to be catalytically active, with average diameters of ~140nm. Using a low-cost metal precursor to induce catalysis could improve process economics over preloading the reactor with metal-impregnated catalytic pellets. Nanoparticle separation and collection on the back end of the reactor is necessary, which adds process complexity.

## 2.10 Process Economics and Performance Metrics

Optimizing the process economics of a continuous SCWG reactor is challenging yet necessary for industrial applications. Competing priorities necessitate elegant and balanced design strategies. Heat recovery is vital to process economics, most commonly accomplished by a heat exchanger used to quench the effluent and preheat the reactor feed. Solid loading >20% is also needed for cost-effective production of gaseous products [3, 11]. However, as noted previously, slow preheating of a premixed reactant feed leads to polymerization and char formation, clogging reactor components and lowering gaseous yields. A regenerative heat exchanger is best used to solely heat feedwater to supercritical conditions, followed by injection of the feedstock. For optimal post-critical injection, a secondary preheater is necessary to heat feedwater well past desired reaction temperatures. Post-critical injection also necessitates pumping the feedstock at much higher solid loading percentage, which is technically challenging [3].

Catalyst integration could theoretically improve process economics, however a detailed techno-economic assessment of catalyst use in SCWG is difficult due to the cost and complexity of catalyst preparation, and uncertainty of catalyst lifetime due to sintering and depletion. Designers must balance material cost of catalyst integration against process improvements, such as lower reaction temperatures, higher viable solid loadings, and higher $H_2$ yields.

Despite the technical challenges noted in this manuscript, there are some factors which are economically favorable for SCWG. Financial incentives exist for effective waste processing, especially for feedstocks such as sewage or toxic chemicals. *In-situ* metal oxide catalyst generation with particle collection could allow for value addition. Finally, direct liquid $CO_2$ capture in the product stream of a reactor could be economically advantageous, as some countries offer tax incentives for carbon oxide capture and sequestration.

From a system perspective, effective SCWG is best described as complete conversion of the mass and energy content of the original feedstock into gaseous products. Three performance metrics are commonly used in the literature to quantify this conversion: (i) gasification efficiency (GE), (ii) carbon conversion efficiency (CE), and (iii) hydrogen efficiency (HE). GE is defined as the ratio of total mass of the gasoues products to initial mass of the feedstock, expressed mathematically as:

$$GE(\%) = \frac{m_{H_2} + m_{CO_2} + m_{CO} + m_{CH_4} + m_{C_xH_y}}{m_{feedstock}} * 100$$

GE is an effective metric for quantifying the overall completeness of gasification reactions. CE is another metric used to quantify completeness of gasification, defined as the ratio of moles of carbon in the gaseous product to moles of carbon in the feedstock:

$$CE(\%) = \frac{2n_{CO_2} + n_{CO} + n_{CH_4} + x\, n_{C_xH_y}}{n_{C,feedstock}} * 100$$

Another, less frequently used, metric is HE, defined as the ratio of moles of hydrogen in the gaseous product to moles of hydrogen in the feedstock:

$$HE(\%) = \frac{2n_{H_2} + 4n_{CH_4} + y\, n_{C_xH_y}}{n_{H,feedstock}} * 100$$

HE and GE values from SCWG can be well above 100%, due to the prominent role of the WGS reaction during gasification. To demonstrate this, the gasification of methanol ($CH_3OH$) is considered, as shown in Figure 4. Methanol first dehydrogenates to form $H_2$ and formaldehyde ($CH_2O$), which subsequently decomposes into $H_2$ and CO. Finally, CO and $H_2O$ are converted to $CO_2$ and $H_2$ via the WGS reaction. 6 mol of H are theoretically present after full gasification, compared to 4 mol of H present in the initial methanol molecule, resulting in a maximum HE of 150%. Similarly the maximum GE is 156%, due to the added molar mass of the water molecule.

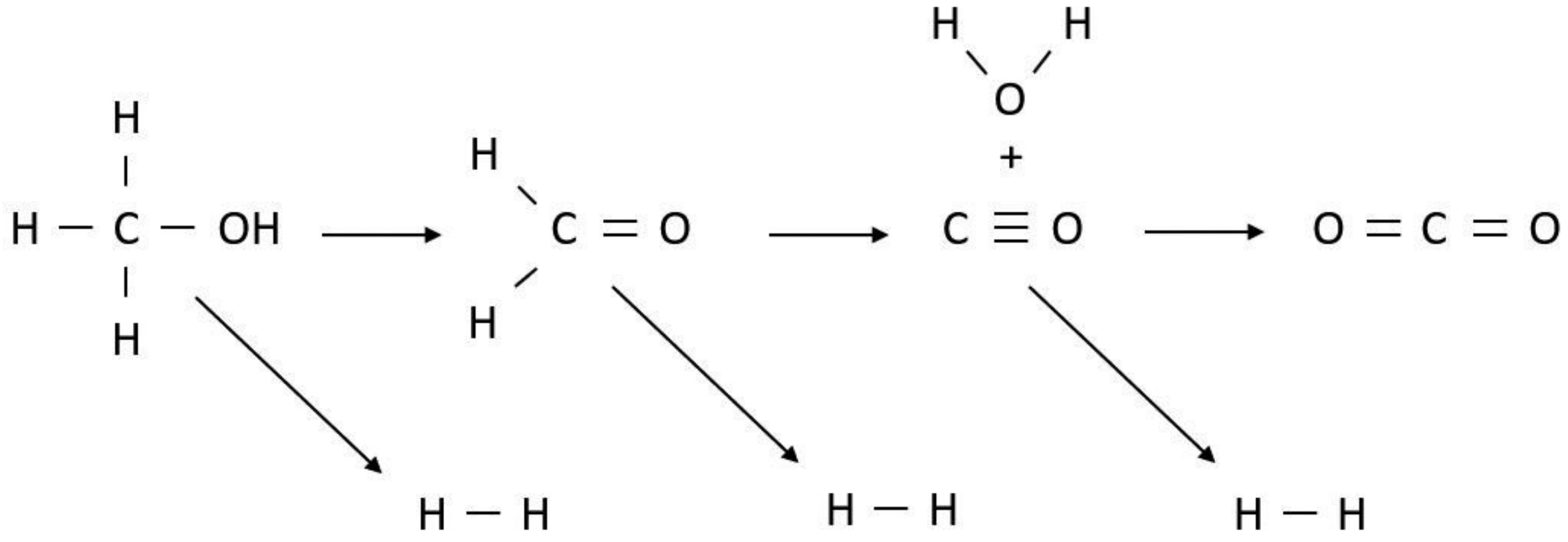

**Figure 4**: Methanol gasification reactions in supercritical water, demonstrating the potential for hydrogen efficiency and gasification efficiency to exceed 100% due to the water-gas shift reaction

## 3. Conclusions

To date, and to our knowledge, no commercial SCWG reactors exist. The reasons for this are twofold: SCWO is more effective for waste destruction, and other methods for producing $H_2$ and $CH_4$ are currently more cost effective. For SCWG to reach commercialization, reactors will need to neutralize waste as efficiently as SCWO, or process economics must be improved. Standardization of reactor designs and identification of solutions to common issues could aid in improving process economics of all SCWG systems.

Studying gasification efficiencies and chemical kinetic rates in lab-scale reactor systems remains an important step in optimizing SCWG reactors. The design of these systems varies with intended use, but the basic requirements of all supercritical water reactor systems reported in the literature are consistent. Pressurization, heating, corrosion mitigation, reactor monitoring and control of process parameters are critical for any system; the solution and strategies for these have been well described in published literature and can be achieved with commercial, off-the-shelf hardware.

Mixing of the reagent and water presents a significant challenge and methodology depends on the design intent of the individual reactors. Laboratory systems will achieve accurate chemical kinetic studies by injecting reagent into supercritical water. Systems processing viscous, heterogeneous feedstocks often have no choice but to premix the water and the reagent prior to heating, which can lead to issues with char formation. For chemical kinetic studies utilizing post-critical mixing, the mixing rate relative to the reaction rate must be considered. Mixing, concentration gradient, and residence time uniformity (assumed in the plug flow scenario) can be enhanced by inducing turbulence or Dean vortices in the reactor section.

Careful evaluation of the mixing process is needed to ensure that the chemical kinetic rate, not the mixing rate, is the limiting rate in studies designed to determine chemical kinetics.

*Ex-situ* analysis methods such as gas chromatography and TOC analysis are not ideal for studying fundamental reaction chemistry. *In-situ* Raman spectroscopic analysis promises to improve monitoring of the gasification process through faster response time and reduced experimental uncertainty. Several designs for optical access to the HTHP section of the reactor have been proposed and tested but maintaining a pressure seal for the optical window remains a challenge.

The promise of recovering fuel value from organic waste in a single-step, environmentally benign process continues to attract attention to supercritical water gasification. To date, the economic hurdle has been significant enough to prevent commercialization, largely due to the technical challenges detailed above. With continued research and development, it may be possible to bring SCWG into widespread commercial use, similar to the emergence of SCWO systems in the industrial sphere.

**Figure Captions**

**Figure 1**: Generalized reaction network for supercritical water gasification of complex organic molecules. Desirable reaction pathways lead to high yields of gaseous products, while undesirable polymerization reactions lead to the formation of char. Heteroatoms or metals present in the feedstock can form metal oxides or inorganic acids, causing significant issues with reactor clogging and corrosion

**Figure 2**: Representative schematic of a continuous supercritical water gasification reactor with premixed water and reagent

**Figure 3**: Plots of baseline-subtracted Raman spectra from formic acid decomposition experiments with (a) Raman probe in cold zone, (b) Raman probe in hot zone at 365°C, (c) Raman probe in hot zone at 380°C, and (d) Raman probe in hot zone at 400°C. Significant reduction in signal intensity appears to be due to temperature-dependent optical effects, which significantly complicate quantitative spectroscopy

**Figure 4**: Methanol gasification reactions in supercritical water, demonstrating the potential for hydrogen efficiency and gasification efficiency to exceed 100% due to the water-gas shift reaction